\documentclass[twocolumn,showpacs,pre,preprintnumbers,amsmath,amssymb]{revtex4}
\usepackage{epsfig}
\usepackage{graphicx}% Include figure files
\usepackage{dcolumn}% Align table columns on decimal point
\usepackage{bm}% bold math
\begin{document}
\title{Evolutionary snowdrift game with loners}
\author{Li-Xin Zhong}
\author{Da-Fang Zheng}\email{dfzheng@zjuem.zju.edu.cn}
\author{B. Zheng}
\affiliation{Zhejiang Institute of Modern Physics and Department
of Physics, Zhejiang University, Hangzhou 310027, People's
Republic of China}
\author{Chen Xu}
\author{P.M. Hui} \affiliation{Department of Physics, The Chinese
University of Hong Kong, Shatin, Hong Kong, China}

\date{\today}

\begin{abstract}
The effects of an additional strategy or character called loner in
the snowdrift game are studied in a well-mixed population or
fully-connected network and in a square lattice.  The snowdrift
game, which is a possible alternative to the prisoner's dilemma
game in studying cooperative phenomena in competing populations,
consists of two types of strategies, C (cooperators) and D
(defectors).  In a fully-connected network, it is found that
either C lives with D or the loners take over the whole
population. In a square lattice, three possible situations are
found: a uniform C-population, C lives with D, and the coexistence
of all three characters.  The presence of loners is found to
enhance cooperation in a square lattice by enhancing the payoff of
cooperators.  The results are discussed in terms of the effects in
restricting a player to compete only with his nearest neighbors in
a square lattice, as opposed to competing with all players in a
fully-connected network.

\end{abstract}

\pacs{87.23.Kg, 02.50.Le, 89.75.Hc, 87.23.Cc}

\maketitle

\section{Introduction}
\label{sec:introduction}

The evolutionary prisoner's dilemma game (PDG)
\cite{axelrod,axelrod1,weibull} and the snowdrift game (SG)
\cite{sugden} have become standard paradigms for studying the
possible emergence of cooperative phenomena in a competitive
setting.  Physicists find such emergent phenomena fascinating, as
similar cooperative effects are also found in interacting systems
in physics that can be described by some minimal models, e.g.
models of interacting spin systems.  These games are also
essential in the understanding of coexistence of (and competition
between) egoistic and altruistic behavior that appear in many
complex systems in biology, sociology and economics.  The basic
PDG \cite{neumann,rapoport} consists of two players deciding
simultaneously whether to cooperate (C) or to defect (D). If one
plays C and the other plays D, the cooperator pays a cost of
$S=-c$ while the defector receives the highest payoff $T=b$
($b>c>0$).  If both play C, each player receives a payoff of
$R=b-c > 0$.  If both play D, the payoff is $P=0$.  Thus, the PDG
is characterized by the ordering of the four payoffs $T>R>P>S$,
with $2R>T+S$. In a single round of the game, it is obvious that
defection is a better action in a fully connected (well-fixed)
population, regardless of the opponents' decisions. Modifications
on the basic PDG are, therefore, proposed in order to induce
cooperations and to explain the wide-spread cooperative behavior
observed in the real world. These modifications include, for
example, the iterated PDG \cite{axelrod,axelrod1}, spatially
extended PDG \cite{nowak,nowak1,doebeli,killingback} and games
with a third strategy \cite{hauert,szabo,szabo1,hauert1}.

The snowdrift game (SG), which is equivalent to the hawk-dove or
chicken game \cite{sugden,smith}, is a model somewhat favorable
for cooperation.  It is best introduced using the following
scenario \cite{hauert2}. Consider two drivers hurrying home in
opposite directions on a road blocked by a snowdrift.  Each driver
has two possible actions -- to shovel the snowdrift (cooperate
(C)) or not to do anything (not-to-cooperate or ``defect" (D)). If
they cooperate, they could be back home earlier and each will get
a reward of $b'$. Shovelling is a laborious job with a total cost
of $c'$.  Thus, each driver gets a net reward of $R=b' - c'/2$. If
both drivers take action D, they both get stuck, and each gets a
reward of $P=0$. If only one driver takes action C and shovels the
snowdrift, then both drivers can also go home. The driver taking
action D (not to shovel) gets home without doing anything and
hence gets a payoff $T=b'$, while the driver taking action C gets
a ``sucker" payoff of $S=b'-c'$.  The SG refers to the case when
$b'>c'>0$, leading to the ranking of the payoffs $T>R>S>P$. This
ordering of the payoffs {\em defines} the SG. Therefore, both the
PDG and SG are defined by a payoff matrix of the form
\begin{equation}
\begin{array}{ccc}
\ & \begin{array}{cc} C & D \end{array} \\
\begin{array}{c} C \\ D \end{array}&\left(\begin{array}{cc}R&S \\
T&P\end{array}\right),
\end{array}
\end{equation}
and they differ only in the ordering of $P$ and $S$.  It is this
difference that makes cooperators persist more easily in the SG
than in the PDG.  In a well-mixed population, cooperators and
detectors coexist.  Due to the difficulty in measuring payoffs and
the ordering of the payoffs accurately in real world situations
where game theory is applicable \cite{milinski,turner}, the SD has
been taken to be a possible alternative to the PDG in studying
emerging cooperative phenomena \cite{hauert2}.

The present work will focus on two aspects of current interest. In
many circumstances, the connections in a competing population are
better modelled by some networks providing limited interactions
than a fully-connected network.  Previous studies showed that
different spatial structures might lead to different behaviors
\cite{nowak,nowak1,hauert3,abramson,ifti}. For example, it has
been demonstrated that spatial structures would promote
cooperation in the PDG \cite{nowak,nowak1}, but would suppress
cooperation in the SG \cite{hauert2}.  There are other variations
on the SG that resulted in improved cooperation
\cite{santos,sysi-ahol}.  Here, we explore the effects of an
underlying network on the evolutionary SG in a population in which
there exists an additional type of players.  The latter is related
to the fact that real-world systems usually consist of people who
would adopt a strategy other than just C and D.  For example,
there may be people who do not like to participate in the
competition and would rather take a small but fixed payoff. Hauert
{\em et al.} studied the effects of the presence of such persons,
called loners \cite{hauert,szabo}, in a generalization of the PDG
called the public goods game(PGG). Motivated by these works of
Hauert {\em et al.} \cite{hauert2,hauert,szabo}, we study the
effects of risk averse loners in the evolutionary SG.  In our
model, evolution or adaptation is built in by allowing players to
replace his character or strategy by that of a better-performing
connected neighbor.  We focus on both the steady state and the
dynamics, and study how an underlying network structure affects
the emergence of cooperation.  It is found that in a
fully-connected network, the C-players and D-players {\em cannot}
coexist with the loners.  In a square lattice, however,
cooperators are easier to survive. Depending on the payoffs, there
are situations in which C-players, D-players and loners can
coexist.

In Sec. \ref{sec:model}, the evolutionary SG with loners in a
population with connections is presented. In Sec.
\ref{sec:Simulation Results and Discussions}, we present detailed
numerical results in fully-connected networks and in square
lattices, and discuss the physics of the observed features.  The
effects of noise are also discussed. We summarize our results in
Sec. \ref{sec:conclusion}.

\section{The Model} \label{sec:model}

We consider an evolutionary snowdrift game in which the
competitions between players are characterized by the payoff
matrix
\begin{equation}
\begin{array}{cccc}
\ & \begin{array}{ccc} C & D & L \end{array} \\
\begin{array}{c} C \\ D \\L \end{array}&\left(\begin{array}{ccc}R&S&Q \\
T&P&Q \\ Q&Q&Q \end{array}\right).
\end{array}
\end{equation}
Here, each player takes on one of three possible characters or
strategies: to cooperate (C), to defect (D), or to act as a loner
(L).  The matrix element gives the payoff to a player using a
strategy listed in the left hand column when the opponent uses a
strategy in the top row.  In the basic SG, it is useful to assign
$R=1$ so that the payoffs can be characterized by a single
parameter $r = c'/2 = c'/(2b'-c')$ representing the cost-to-reward
ratio \cite{hauert2}. In terms of $0< r<1$, we have $T=1+r$,
$R=1$, $S=1-r$, and $P=0$. A competition involving a loner leads
to a payoff $Q$ for both players.  Here, we explore the range of
$0<Q<1$.

Spatial networking effects and evolutions are incorporated into
the SG as follows.  At the beginning of the game, the players are
arranged onto the nodes of a network and the character $s(i)$ of
each player is assigned randomly among the choices of C, D, and L.
Our discussion will be mainly on fully-connected graphs and
regular lattices.  In a fully-connected network, every player is
connected to all other players.  In a square lattice, a player is
linked only to his four nearest neighbors.  Numerical studies are
carried out using Monte Carlo simulations as reported in the work
of Szab\'{o} {\em et al.} \cite{szabo1} (see also Refs.
\cite{szabo,hauert1}). The evolution of the character of the
players is governed by the following dynamics. At any time during
the game, each player competes with all the players that he is
linked to and hence has a payoff.  A randomly chosen player $i$
reassesses his own strategy by comparing his payoff $P(i)$ with
the payoff $P(j)$ of a randomly chosen connected neighbor $j$.
With probability
\begin{equation}
W[s(i),s(j)] = \frac{1}{1 + \exp\left([P(i) - P(j)]/K\right)},
\end{equation}
the player $i$ adopts the strategy of player $j$.  Otherwise, the
strategy of player $i$ remains unchanged.  Here $K$ is a noise
parameter \cite{szabo,szabo1,hauert1} that determines the
likelihood that player $i$ replaces his strategy when he meets
someone with a higher payoff. For $K \approx 0$, a player $i$ is
almost certain to replace (not to replace) his strategy when he
meets someone with a better (worse) payoff.  For large $K$, a
player has a probability of $1/2$ to replace his strategy,
regardless of whether $P(j)$ is better or worse than $P(i)$.  In a
fully connected network, a player's character may be replaced by
any player in the system. In a square lattice, a player's
character can only be replaced by one of his four connected
neighbors.  As the game evolves, the fractions of players with the
three characters also evolve. These fractions are referred to as
frequencies. Depending on the parameters $r$ and $Q$, the
cooperator frequency $f_{C}$, defector frequency $f_{D}$, and
loner frequency $f_{L}$ take on different values in the long time
limit.

\section{Results and Discussions}
\label{sec:Simulation Results and Discussions}

\subsection{Fully-connected network}

\begin{figure}
\begin{center}
\epsfig{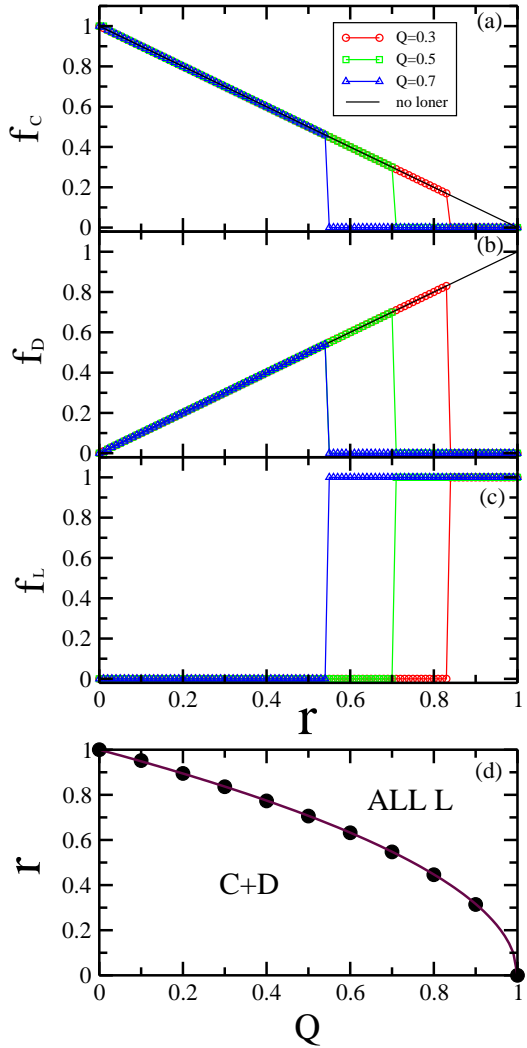}
\caption{\small{(Color online) (a) Cooperator frequency $f_{C}$,
(b) defector frequency $f_{D}$, (c) loner frequency $f_{L}$ as a
function of $r$ for three different values of the loner payoff
$Q=0.3$, $0.5$, and $0.7$ in a fully-connected network. The
results of the snowdrift game without loners are also included for
comparison (solid lines). (d) Phase diagram showing the two phases
separated by $r_{L}(Q)$ in the $r$-$Q$ parameter space.  The
symbols show the numerical results of $r_{L}(Q)$ and the line
gives the functional form $\sqrt{1-Q}$.}}
\end{center}
\end{figure}

We performed detailed numerical studies on our model.  The number
of players in the system is taken to be $N=10^{4}$.  In getting
the fraction of players of different characters in the long time
limit, we typically average over the results of $10^{3}$
Monte-Carlo time steps per site (MCS), after allowing $5 \times
10^{3}$ MCS for the system to reach the long time limit. Averages
are also taken over 10 initial configurations of the same set of
parameters.  Figure 1 shows the results for fully connected
networks.  A value of $K=0.1$ is taken. The cooperator frequency
$f_{C}$, defector frequency $f_{D}$, and loner frequency $f_{L}$
are obtained as a function of the cost-to-benefit ratio $r$ for
three different values of the loner's payoff $Q=0.3$, $0.5$, and
$0.7$. In the absence of loners \cite{hauert2}, $f_{C}(r) = 1-r$
and $f_{D}=r$ in a fully connected network. From Figure 1, the
loners extinct for a range of values of $r < r_{L}(Q)$ in which
the behavior is identical to the basic SG.  For $r > r_{L}(Q)$,
the loners invade the whole population and both cooperators and
defectors disappear. This is similar to the results in the PDG
\cite{szabo1} and in the PGG \cite{hauert}. In a fully connected
network, the three characters {\em cannot} coexist. This is in
sharp contrast to the rock-scissors-paper game
\cite{tainaka,frean,szolnoki} on a fully connected network in
which the three strategies coexist.  We obtained $r_{L}(Q)$
numerically.  The result is shown in Figure 1(d) as a curve in the
$r$-$Q$ parameter space.  It is found that $r_{L}(Q)$ follows the
functional form $\sqrt{1-Q}$, which will be explained later. The
curve $r_{L}(Q)$ represents a phase boundary that separates the
$r$-$Q$ space into two regions. The region below (above) the curve
corresponds to a phase in which cooperators and defectors (only
loners) coexist (exist).

\begin{figure}
\begin{center}
\epsfig{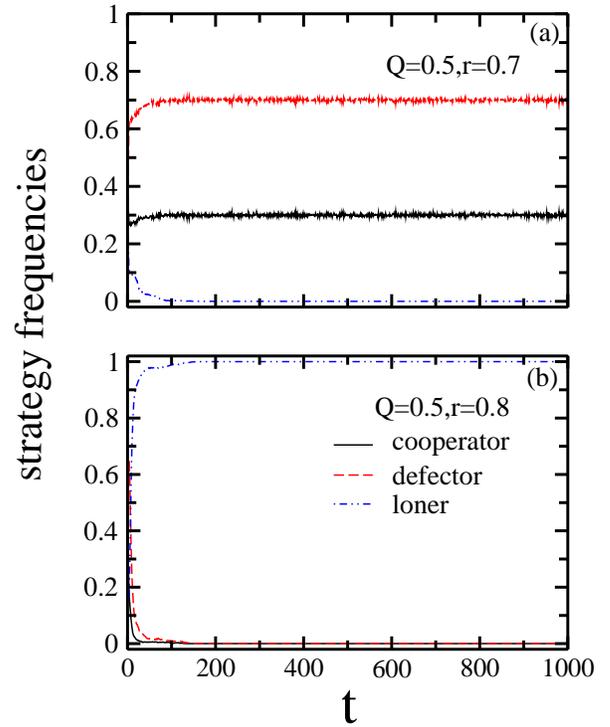}
\caption{\small{(Color online) Typical short-time behavior of the
cooperator, defector, and loner frequencies in a fully-connected
network for (a) the phase that C-players and D-players coexist
($Q=0.5$ and $r=0.7$); and (b) the phase that loners take out the
whole system ($Q=0.5$ and $r=0.8$).}}
\end{center}
\end{figure}

We also studied the temporal evolution in both phases, i.e., for
$r < r_{L}(Q)$ and $r > r_{L}(Q)$.  Taking $Q=0.5$, for example,
$r_{L} = 1/\sqrt{2} = 0.707$.  Figure 2 shows $f_{C}(t)$,
$f_{D}(t)$ and $f_{L}(t)$ in the first $10^{3}$ MCS.  The initial
frequencies are $1/3$ for all three characters.  For values of $r$
deep into either phase (see Fig. 2), the transient behavior dies
off rapidly and the extinct character typically vanishes after
$\sim 10^{2}$ MCS.  In the phase where C and D coexist, $f_{C}$
and $f_{D}$ oscillate slightly with time in the long time limit,
due to the dynamical nature of the game.  It is noted that for $r
\approx r_{L}$, the strategies compete for a long while and the
transient behavior lasts for a long time. This slowing down
behavior is typical of that near a transition.

The behavior of $r_{L}(Q) = \sqrt{1-Q}$ follows from the rule of
character evolution.  In a fully-connected network, all C-players
have the {\em same} payoff $P(C)$ and all D-players have the {\em
same} payoff $P(D)$.  These payoffs depend on $f_{C}$, $f_{D}$,
and $f_{L}$ at each time step.  The payoff for a loner is $NQ$ at
all time, for a system with $N \gg 1$.  For small $K$, $f_{L}$
decays exponentially with time if $P(C)$ and $P(D)$ are both
greater than $NQ$. In addition, the phase with only non-vanishing
$f_{C}$ and $f_{D}$ is achieved by having $P(C)=P(D)$.  For this
phase in the long time limit, $P(C) = N(f_{C} + f_{D}(1-r))$ and
$P(D) = N f_{C}(1+r)$. Together with $f_{C}+f_{D}=1$ (since
$f_{L}=0$ in the phase under consideration), the condition
$P(C)=P(D)$ implies $f_{C} = 1-r$ and $f_{D}=r$. These results are
identical to the basic SG (without loners) in a fully connected
network.  The validity of this solution requires $P(C)>NQ$ (and
hence $P(D) > NQ$), which is equivalent to $r < \sqrt{1-Q}$.  This
is exactly the phase boundary shown in Figure 1(d).

\subsection{Square Lattice}

\begin{figure}
\begin{center}
\epsfig{figure=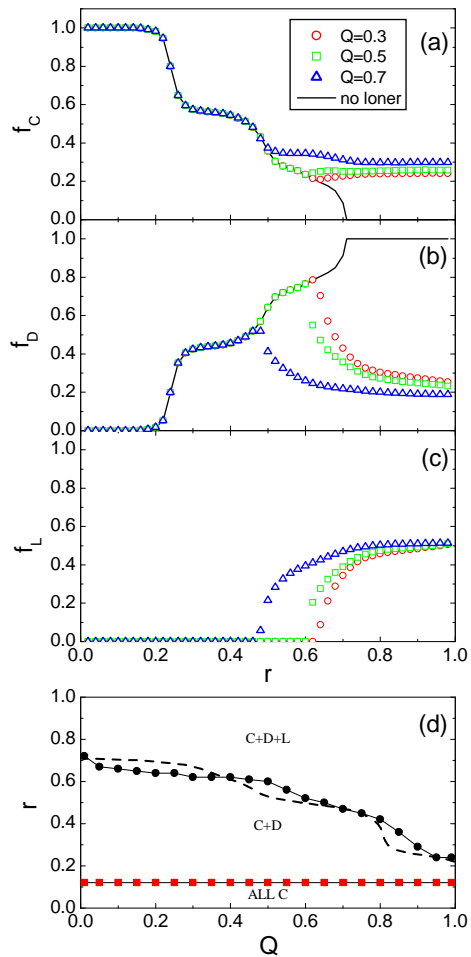,width=1.2\linewidth}
\caption{\small{(Color online) (a) Cooperator frequency $f_{C}$,
(b) defector frequency $f_{D}$, (c) loner frequency $f_{L}$ as a
function of $r$ for three different values of the loner payoff
$Q=0.3$, $0.5$, and $0.7$ in a square lattice. The results of the
snowdrift game without loners in a square lattice are also
included for comparison (solid lines). (d) Phase diagram showing
the different phases in the $r$-$Q$ parameter space.  The dashed
line shows the phase boundary obtained by an approximation as
discussed in the text.}}
\end{center}
\end{figure}

\begin{figure}
\begin{center}
\epsfig{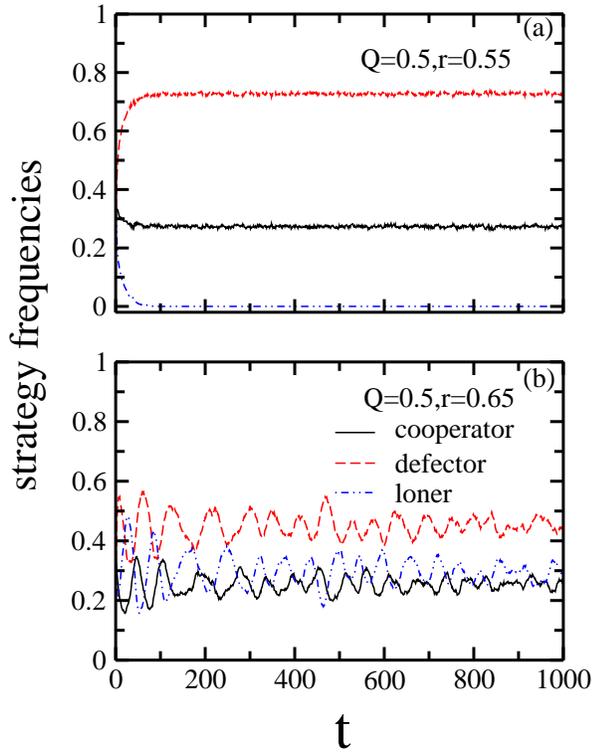}
\caption{\small{(Color online) Typical short-time behavior of the
cooperator, defector, and loner frequencies in a square lattice
for (a) the phase that C-players and D-players coexist ($Q=0.5$
and $r=0.55$); and (b) the phase that C-players, D-players, and
loners coexist ($Q=0.5$ and $r=0.65$).}}
\end{center}
\end{figure}

The behavior of the game in a square lattice is expected to be
quite different, due to the restriction that a player can only
compete with his connected neighbors.  We carried out simulations
on $100 \times 100$ square lattices with periodic boundary
conditions.  Figure 3(a)-(c) shows $f_{C}(r)$, $f_{D}(r)$ and
$f_{L}(r)$ for three different values of the loner payoff $Q$. The
results for the spatial SG (without loners) on a square lattice
\cite{hauert2} is also shown (solid lines in Figure 3(a) and 3(b))
for comparison.  A value $K=0.1$ is used.  Several features should
be noted.  For $r<r_{L}^{(SL)}(Q)$, the loners eventually vanish
with $f_{C}$ and $f_{D}$ take on the mean values in the spatial SG
without loners. This behavior is similar to that in fully
connected networks.  For $r > r_{L}^{(SL)}(Q)$, however, the
behavior is different from that in fully connected networks. Here,
C, D, and L characters coexist.  Above $r_{L}^{(SL)}$, $f_{D}$
drops with $r$ to a finite value, leaving rooms for $f_{L}$ to
increase with $r$. The cooperator frequency remains finite above
$r_{L}^{(SL)}$.  Therefore, the cooperator frequency or the
cooperative level in the system as a whole is significantly {\em
improved} by the presence of loners.  For $r > r_{L}^{(SL)}$,
increasing the payoff $Q$ of loners leads to a higher cooperator
frequency and lower defector frequency. Reading out $r_{L}^{(SL)}$
for different values of $Q$, we get the phase boundary as shown in
Figure 3(d) that separates a region characterized by the
coexistence of three characters and a region in which only C and D
coexist. The results indicate that, due to the restriction imposed
by the spatial geometry that a player can only interact with his
four nearest neighbors, it takes a certain non-vanishing value of
$r$ for loners to survive even in the limit of $Q \rightarrow 1$.
The behavior is therefore different from that in a fully connected
network for which the boundary is given by $\sqrt{1-Q}$. Note that
there exists a region of small values of $r$ in which the steady
state consists of a uniform population of C strategy (see Fig.
3(a) and Fig. 3(d)).  For small $Q$, loners are easier to survive,
when compared with the fully connected case. Putting these results
together, the phase diagram (see Fig. 3(d)) for a square lattice,
therefore, shows three different phases.  The most striking effect
of the spatial structure is that cooperators now exist in every
phase.

Interestingly, we found that the phase boundary $r_{L}^{(SL)}(Q)$
in Figure 3(d) can be described quantitatively as follows.  We
{\em assume} that the survival of loners is related to the
cooperator frequency.  In particular, loner survival requires the
cooperator frequency to drop below a certain level
$\overline{f}(Q)$ and that this value is the same in a square
lattice as in a fully connected network.  That is to say, we
assume that loners could survive, for a given value of $Q$ and
$K$, only when $f_{C} < \overline{f}(Q) = 1 - \sqrt{1-Q}$.
Numerical results also indicate that when all loners extinct,
$f_{C}$ and $f_{D}$ follow the results in a spatial SG without
loners.  This is shown as the solid line in Figure 3(a).
Therefore, for a given value of $Q$, we can simply read out the
value of $r$ such that $f_{C} = \overline{f}(Q)$ from the results
in spatial SG in a square lattice. For different values of $Q$,
this procedure results in the dashed line shown in Figure 3(d)
which describes the phase boundary quite well.

Figure 4 shows the temporal dependence of $f_{C}$, $f_{D}$, and
$f_{L}$ in a square lattice for two values of $r$ at $Q=0.5$. For
$r=0.55$ (Fig. 4(a)), which corresponds to a case in which only
cooperators and defectors coexist, the number of loners decay
rapidly in time, typically within 100 MCS. After the transient
behavior, the cooperator and defector frequencies only oscillate
slightly about their mean values.  This behavior is similar to
that in the C and D coexistence phase in Figure 1(d) for
fully-connected networks.  For $r=0.65$ (Fig. 4(b)), which
corresponds to a case with the three characters coexist, the long
time behavior of $f_{C}$, $f_{D}$ and $f_{L}$ is oscillatory.
Similar behavior has been found in the rock-scissors-paper game
\cite{tainaka,frean,szolnoki} and in the voluntary PDG
\cite{hauert1}. Due to the dynamical nature of character
evolution, there are continuous replacements of one character by
another and this oscillatory behavior is expected.

\begin{figure}
\begin{center}
\epsfig{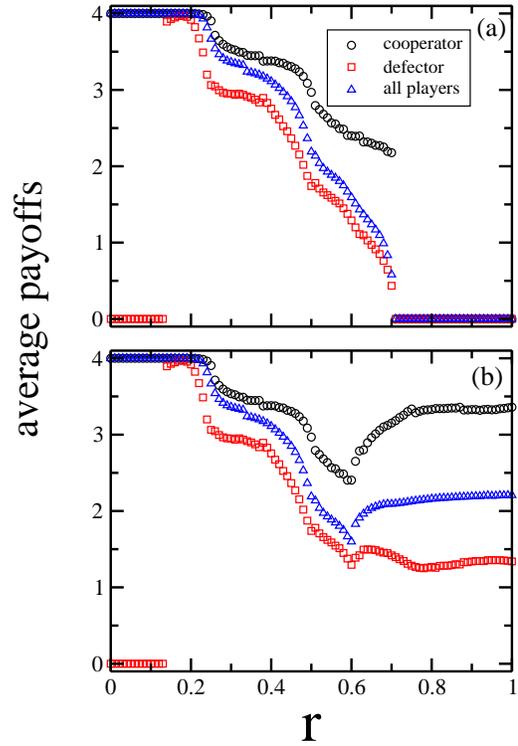}
\caption{\small{(Color online) (a) Average payoffs of each
character as a function of $r$ in a snowdrift game without loners
on a square lattice.  The payoff averaged over all players is also
shown.  (b) Average payoffs of cooperators and defectors as a
function of $r$ in a snowdrift game with loners. The parameters
are $Q=0.5$ and $K=0.1$.  Note that the loners, if exist, have a
constant payoff of $4Q$. The payoff averaged over all players is
also shown.}}
\end{center}
\end{figure}

The major difference between a square lattice and a
fully-connected network is that in a fully-connected network, each
player competes with all other players.  As a result, there are
only three payoffs in the system -- one for each type of player,
at each time step. The loners, for example, have a constant payoff
of $NQ$, while the cooperators and defectors have payoffs that
depend on $f_{C}(t)$ and $f_{D}(t)$.  Once $NQ$ is higher than the
payoffs of cooperators and defectors, the number of loners grows
until they take over the whole population.  In a square lattice,
however, each player has a payoff that depends on his character
{\em and} the detail of his neighborhood, i.e., the characters of
his four connected neighbors.  This implies that the C-players and
D-players in a square lattice may have different payoffs depending
on the characters of his connected neighbors. The loners have a
constant payoff of $4Q$.  The non-uniform payoffs among C-players
and D-players in a lattice allow some C and D players to coexist
with the loners, by evolving to spatial local configurations that
favor their survivals.

Since the adaptive rule is related to the payoff of each
character, it will be interesting to compare the payoffs in a
spatial SG without and with loners.  Figure 5(a) shows the mean
payoffs of cooperators and defectors as a function of $r$ in a
spatial SG in a square lattice {\em without} loners.  The averaged
payoff over all players is also shown.  For small $r$, there is a
phase with all C players and the payoff is 4 for each of the C
players. For large $r$, there is a phase with all D players and
the payoff is zero. For intermediate $r$ where C and D players
coexist, the mean payoff drops gradually with $r$.  In a spatial
SG {\em with} loners (Fig. 5(b)), it is observed that the mean
payoffs basically follow that in Figure 5(a) in the phase where
loners are completely replaced.  When loners can survive, the
presence of these loners increases the payoffs of both the
remaining cooperators and defectors.  The loners themselves have a
payoff of $2$ in a 2D square lattice.  The cooperators' payoff is
enhanced once loners survive and the increase follows the same
form as the increase in the loner frequency with $r$ (compare the
circles in Fig. 5(b) with the squares in Fig. 3(c) in the range of
$r$ when loners survive). When loners survive, the payoff averaged
over all players is significantly {\em enhanced} due to their
presence. This is similar to what was found in the voluntary PDG
\cite{hauert1}.

\subsection{Effects of noise}

All the results reported so far are for the case of $K=0.1$.  This
corresponds to a case where the player is highly likely to replace
his character when he meets a better-performing player.   In
Figure 6, we show the effects of the noise parameter for a fixed
$Q=0.3$. As $K$ increases, the step-like structure in $f_{C}$ as a
function of $r$ becomes less obvious and $f_{C}$ is gradually
suppressed in the $r \rightarrow 1$ limit.  The most important
effect of a 2D square lattice is that each player is restricted to
interact with his four neighbors.  Take a player of character
$s(i)$, he will only encounter a finite number of configurations
for which he is competing in.  For example, his four neighbors may
consist of 4 C-players; 3 C-players and 1 D-player or 1-loner,
etc.  Each of these configurations corresponds to a $P(i)$.  In a
square lattice, therefore, there will be a finite number of
payoffs for a C-player, depending on the characters of the
neighbors. Similarly, there are finite number of payoffs for a
D-player. The loners always get a payoff of $4Q$. For $K \approx
0$, the adaptive mechanism is strictly governed by the ordering of
these payoffs.  The distribution of players in a square lattice
will then evolve in time according to how the payoffs are ordered.
In the long time limit, only a few favorable local configurations
will survive and the number of players in each of these favorable
configurations is high.  As one increases $r$ slightly, the
ordering of the finite number of payoffs may not change.
Therefore, $f_{C}$ will not change with $r$ until we reach certain
values of $r$ that the ordering of the payoffs is changed. This
gives rise to the more sudden changes in $f_{C}$ as observed at
some values of $r$ and it is the reason for having step-like
features in $f_{C}$ and $f_{D}$ for small values of $K$. As the
noise parameter $K$ increases, the adaptive mechanism is less
dependent on the exact ordering of the payoffs. Therefore, the
changes in $f_{C}$ with $r$ becomes more gradual as $K$ increases.
Interestingly, less obvious step-like structures in $f_{C}$ are
also observed in the spatial SG without loners in 2D lattices with
a larger coordination number \cite{hauert2}.  This is also related
to the picture we just described.  A lattice with more neighbors
will give a higher number of neighborhood configurations and hence
more values of the payoffs. More configurations also imply the
number of players encountering a certain configuration is smaller.
Thus, the number of players involved in a change in the ordering
of the payoffs as $r$ changes is smaller.  This has the effect of
making the drop in $f_{C}$ gradual.  Therefore, increasing $K$ for
a given fixed coordination number is similar in effect as
increasing the coordination number for fixed $K$.

\begin{figure}
\begin{center}
\epsfig{figure=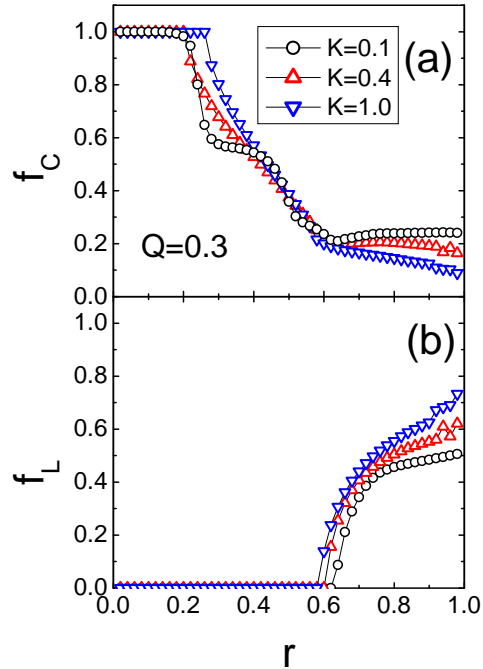,width=0.9\linewidth}
\caption{\small{(Color online) (a) The cooperator frequency
$f_{C}$ and (b) the loner frequency $f_{L}$ as a function of $r$
for three different values of the noise parameter $K=0.1$, $0.4$
and $1.0$.}}
\end{center}
\end{figure}

\section{Summary}
\label{sec:conclusion}

We studied the effects of the presence of loners in a snowdrift
game with loners in fully-connected networks and in square
lattices. In a fully-connected network, either cooperators live
with defectors or loners take over the whole population.  The
condition for loners to take over is found to be $Q > 1-r^{2}$.
This result can be understood by following the payoffs of each
strategy.  In a fully-connected network, the strategies' payoffs
are particularly simple in that they depend only on the strategy
frequencies at the time under consideration, with each type of
player having the same payoff.

In a square lattice, the spatial SG with loners behave quite
differently.  It is found that the cooperators can survive in the
fully parameter space covering $0<r<1$ and $0<Q<1$.  Depending on
the values of these parameters, there are three possible phases: a
uniform C-player population, C-players and D-players coexist, and
coexistence of the three characters.  The underlying lattice thus
makes the survival of cooperators easier.  The presence of loners
is also found to promote the presence of cooperators.  There
average payoff among all players is also found to be enhanced in
the presence of loners.  We discussed the influence of a square
lattice in terms of the payoffs of the players.  In a square
lattice, spatial restriction is imposed on the players in that a
player can only interact with the four nearest neighbors.  This
leads to a payoff that does not only depend on the character but
also depend on the local environment in which the player is
competing in.  The players in the local environment, in turns, are
also competing in their own local environment.  This will lead to
clustering or aggregation of players in the square lattice into
configurations that the payoffs favored.  The dependence of the
frequencies on $r$ in a square lattice then reflects the change in
preferred configurations as $r$ is changed.

We also studied the effects of the noise parameter in the adaptive
mechanism.  It is found that as the noise parameter increases, the
change of the frequencies with $r$ becomes more gradual.  This is
related to the importance of the ordering of the many payoffs in
the adaptive mechanism.  As the noise parameter increases, the
exact ordering of the payoffs becomes less important and the
change in frequencies becomes more gradual.

In closing, we note that it will be interesting to further
investigate the effects of loners in the snowdrift game in
networks of other structures.  Among them are the re-wiring of
regular lattices into a small-world network or a random network
and the scale-free networks \cite{albert}.

\section*{ACKOWLEDGMENTS}
This work was supported in part by the National Natural Science
Foundation of China under Grant Nos. 70471081, 70371069, and
10325520, and by the Scientific Research Foundation for the
Returned Overseas Chinese Scholars, State Education Ministry of
China. One of us (P.M.H.) acknowledges the support from the
Research Grants Council of the Hong Kong SAR Government under
Grant No. CUHK-401005 and from a Direct Research Grant at CUHK.

%************************************************************

\end{document}